\documentclass[conference]{IEEEtran}
\IEEEoverridecommandlockouts
\usepackage{cite}
\usepackage{amsmath,amssymb,amsfonts}
\usepackage{algorithmic}
\usepackage{graphicx}
\usepackage{textcomp}
\usepackage{xcolor}
\usepackage{comment}
\def\BibTeX{{\rm B\kern-.05em{\sc i\kern-.025em b}\kern-.08em
    T\kern-.1667em\lower.7ex\hbox{E}\kern-.125emX}}
\begin{document}

\title{An extended reality-based framework for user risk training in urban built environment\\
\thanks{This work was supported by the CLIMRES “Leadership for Climate Resilient Buildings” Horizon Innovation Actions project supported by the European Commission under grant agreement No 101147777. 
}
}

\author{\IEEEauthorblockN{1\textsuperscript{st} Sotirios Konstantakos}
\IEEEauthorblockA{\textit{Department of Research and Development} \\
\textit{MOBICS}\\
Athens, Greece \\
skon@mobics.gr}
\and
\IEEEauthorblockN{2\textsuperscript{nd} Sotirios Asparagkathos}
\IEEEauthorblockA{\textit{R\&D and Innovation Department} \\
\textit{SingularLogic}\\
Athens, Greece \\
saspragkathos@singularlogic.eu}
\and
\IEEEauthorblockN{3\textsuperscript{nd} Moatasim Mahmoud}
\IEEEauthorblockA{\textit{R\&D and Innovation Department} \\
\textit{SingularLogic}\\
Athens, Greece \\
mmoatasim@singularlogic.eu}
\and
\IEEEauthorblockN{4\textsuperscript{th} Stamatia Rizou}
\IEEEauthorblockA{\textit{R\&D and Innovation Department} \\
\textit{SingularLogic}\\
Athens, Greece \\
srizou@singularlogic.eu}
\and
\IEEEauthorblockN{5\textsuperscript{th}Enrico Quagliarini}
\IEEEauthorblockA{\textit{DICEA} \\
\textit{Polytehcnical University of Marche}\\
Ancona, Italy \\
e.quagliarini@staff.univpm.it}
\and
\IEEEauthorblockN{6\textsuperscript{th}Gabriele Bernardini}
\IEEEauthorblockA{\textit{DICEA} \\
\textit{Polytehcnical University of Marche}\\
Ancona, Italy \\
g.bernardini@univpm.it}

}

\maketitle

\begin{abstract}
In the context of increasing urban risks, particularly from climate change-induced flooding, this paper presents an extended Reality (XR)-based framework to improve user risk training within urban built environments. The framework is designed to improve risk awareness and preparedness among various stakeholders, including citizens, local authorities, and emergency responders. Using immersive XR technologies, the training experience simulates real-world emergency scenarios, contributing to active participation and a deeper understanding of potential hazards and especially for floods. The framework highlights the importance of stakeholder participation in its development, ensuring that training modules are customized to address the specific needs of different user groups. The iterative approach of the framework supports ongoing refinement through user feedback and performance data, thus improving the overall effectiveness of risk training initiatives. This work outlines the methodological phases involved in the framework's implementation, including i)user flow mapping, ii)scenario selection, and iii)performance evaluation, with a focus on the pilot application in Senigallia, Italy. The findings underscore the potential of XR technologies to transform urban risk training, promoting a culture of preparedness and resilience against urban hazards. 
\end{abstract}


\section{Introduction}

Resilience in urban built environments is fundamentally supported by structural and non-structural solutions that improve their capacity to withstand and recover from hazards such as natural disasters and climate change impacts \cite{trainingcampaing}. Structural solutions encompass engineering practices, including the design and retrofitting of buildings and open spaces to improve their durability and strength, and can be supported by urban planning actions. 
The joint implementation of these strategies can contribute to support cities in preparing for and responding to various hazards, ultimately safeguarding lives and property.

Among the non-structural solutions, improving the awareness of end-users towards possible risk conditions and making them prepared to autonomously face them in a safe and proper manner, represents one of the fundamental strategies to improve the resilience of the urban built environment, as also underlined by the Sendai framework \cite{Sendai}. End-user training activities should be designed for citizens, local authorities, rescuers, and designers, by comprising different typologies by individual vulnerability (such as those related to sensory, motion, gender, age, social and cultural issues) and grouping end-users depending on their role as stakeholders in built environment resilience \cite{vrcalvaresi,trainingcampaing}. 

In this perspective, approaches using Virtual and Augmented Reality, and, more in general, of Extended Reality (XR), \cite{XRdefinition,XRdefinition2}, can boost risk awareness and emergency response \cite{trainingscorgie}. Since XR can be defined as “an environment containing real or virtual components or a combination thereof, where the variable X serves as a placeholder for any form of new environment” \cite{XRdefinition2}, applications to safety training can balance common issues for learning support with customization of contents and interactions in emergency scenarios \cite{vrcalvaresi}. Moreover, XR solutions enable the creation of multiple simulated conditions, avoiding complexities of real-world drills. XR can contribute to an increase in the end-users' engagement in training experiences, thus generally providing higher levels of knowledge acquisition and retention with respect to traditional training activities \cite{trainingscorgie}. Combining XR and Serious gaming techniques significantly supports these goals, creating "immersive and stimulating environments for users" \cite{SGdefinition}, who are asked to solve tasks and then receive feedback from a "learn-by-doing" perspective without exposing users to risks. 

XR solutions for end-user training have been widely applied to the scale of single buildings, involving occupants and first responders with respect to hazard prevention, safety behaviours, emergency response, and evacuation \cite{vrcalvaresi}. These solutions are mainly devoted to indoor contexts, and linked to fire safety, but additional applications to earthquake and flood risk have been provided \cite{vrfloodriskperception,trainingfire,trainingscorgie}. A more limited number of research prototypes are also now available in the context of outdoor and urban scenarios \cite{vrflooddamico,vr360lovreglio,warningvirtual}. Nevertheless, many relevant risks affect  urban areas and thus wide-scale training campaigns should be encouraged to improve risk awareness and preparedness of citizens and emergency stakeholders \cite{Sendai}. Efforts to define, implement and validate integrated frameworks for end-users' training against risks at the urban scale should be encouraged. Multi-role and integrated platforms for training actions towards  citizens (as final "user-at-risk"),  first responders,   designers of mitigation strategies are needed, ensuring the proper coordination at the urban level under the supervision of local administrations as central entities for risk reduction, capacity building, emergency communication and management \cite{municipaldisaster}.

This work aims at defining basic criteria for an integrated, user-centred XR-based framework for risk awareness, preparedness and training, considering the application to risks emerging in urban scenarios. The framework development includes the direct involvement of stakeholders, i.e. local administrations, to derive the XR-based requirements. This kind of action implies the selection of pilots for preliminary definition and verification. It also will be set up to ensure replication, scalability and sustainability of the proposed solutions, thus taking into account strategies for the generalization of pilot-based assumptions to other contexts.

In greater detail, this work is part of the European project CLIMRES - "Leadership for climate resilient buildings" (www.climres.eu). CLIMRES aims to provide practical advice and tools to assist stakeholders engaged in combating climate change in evaluating and planning effective interventions for risk reduction and mitigation, in which those related to end-users' preparedness assume a paramount role in boosting the whole urban resilience level. 

\section{PHASES AND METHODS}
\subsection{Phases}
The work is organized in three phases. The first phase concerns the definition of the general user flow in a workshop with project partners (Section II.B). The second phase concerns the definition of user stories and requirements by end-users, i.e. the local authorities of selected pilots (Section II.C). The final phase concerns the integrated workflow definition, which comprises the pilot requests (Section II.D).  

\subsection{User flow definition methods}

The user flow definition first implies   understanding  the functionalities needed to meet user needs, thanks to literature works analysis (Section I)  and collaborative sessions with stakeholders and all CLIMRES partners.
In particular, stakeholder engagement in the design process helps ensure that the outputs are aligned with user expectations and framework goals, guiding the transition from conceptualization to tangible results such as wireframes and mock-ups.

In practical terms, these collaborative sessions  foster open dialogue and collective brainstorming, and aim to map user flows, serving as a blueprint for user interactions within the framework and thus the application service provided by CLIMRES project as final output. In this sense, the identification of target users for the framework were first provided, thus defining the starting point of the flow.


Then, motivations behind user type engagement with the workflow were explored,  discussing the goals to achieve, through specific user flows.
To maintain clarity and focus, the user flow was limited to 10 to 20 high-level actions. This constraint also facilitates easier interpretation and implementation in the design phase. Thus, common features across flows and unique functionalities specific to certain applications was traced.

\begin{figure}[b]
    \centering
    \includegraphics[width=\linewidth]{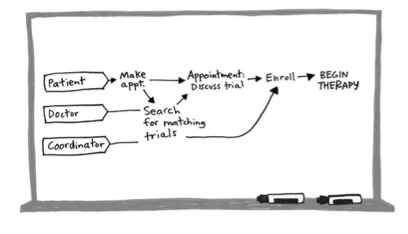}
    \caption{Schematic approach of user flow exercise }
    \label{fig:user_flow_def}
\end{figure}

Fig. \ref{fig:user_flow_def}, illustrates an example of the overall final flow exercises with stakeholders, showcasing the logical progression users would follow from initiation to goal completion.

The identification of common and unique features through user flow mapping is instrumental in aligning design efforts with user needs and service objectives. Engaging stakeholders in the design process ensures that the resultant wireframes and mock-ups are user-centric and functional. This collaborative approach mitigated the risk of misalignment between user expectations and service delivery, facilitated a shared understanding of user needs and fostered a sense of ownership among participants.

\subsection{User stories and requirements methods for final framework definition}

The CLIMRES project employs a user-centred design approach to gather and analyze user-specific needs, challenges, and requirements. In this perspective, additional workshops with pilots have been considered in the framework development, to collect their insights through user stories in view of further application to their specific urban contexts. This collaborative and iterative process, including reviews and follow-up activities, refines user stories and aligns them with project goals. 

To this end, a structured template was distributed to pilot partners, guiding stakeholders to create a user story that includes a description of the main objectives, a list of specific requirements, and an assessment of the data availability necessary for implementation. This structured approach allows for a clear understanding of user needs. This approach ensures the framework alignment with stakeholder expectations and technical realities.

Results are organized by first showing the general user flow diagram, integrated with the feedback from the pilot context, as the result of the workshop with the pilots carried out from November 2024 to January 2025. Then, the typology of training platform selected by the pilot is discussed. Then, specific details on the steps for the XR-based framework are proposed on these bases. Finally, specifications of each element in the framework are proposed to provide fundamentals of the operational phases, which will be developed in the next project steps, along with suggestions on how to measure the performance of the training solutions.




Then, user stories and requirements were used to create a preliminary set of technical requirements, which involved technical partners actively creating the final operational services starting from the framework. A dynamic and responsive approach was used indeed, assessing each requirement for feasibility based on user requirements from pilots, and providing mitigation strategies  to reframe unfeasible ones. This approach ensures the framework alignment with stakeholder expectations and technical realities.




\subsection{Pilot description}
\begin{figure}
    \centering
    \includegraphics[width=1\linewidth]{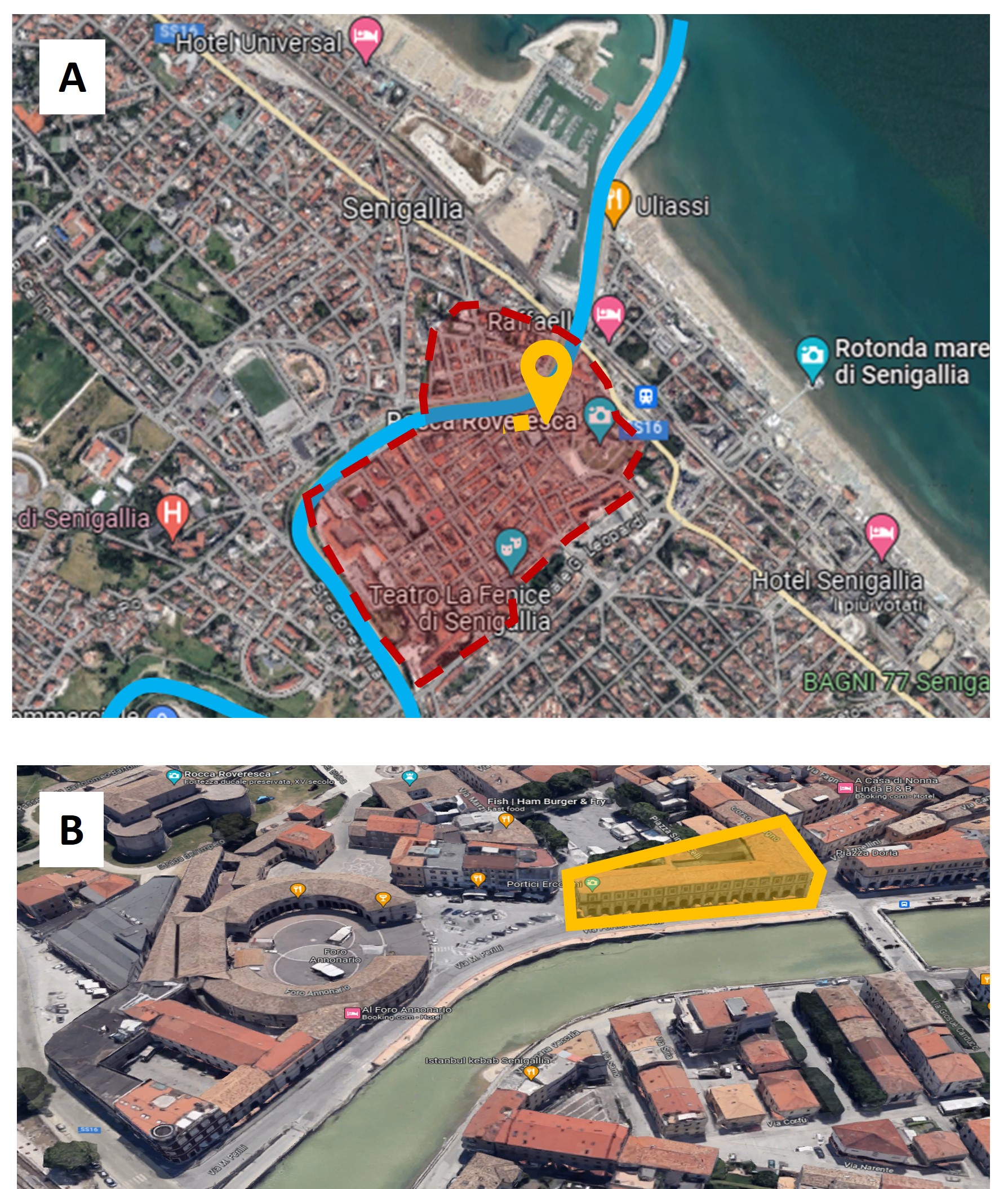}
    \caption{Senigallia overview as the main CLIMRES pilot for the XR-based framework application: A-plan view of the city centre, marking the historical city center (in red), the Misa River (blue line) and the main application area and pilot building (yellow marker); B- aerial view of the main application area and the pilot building (in yellow). (Base map from Image ©2024 Google, Image ©2023 Airbus, Maxar Technologies, Dati cartografici ©2024). }
    \label{fig:pilot_overview}
\end{figure}

The city of Senigallia (AN, Italy) has been selected as a relevant demonstration pilot for the XR-based framework, being involved in the CLIMRES project in view of its relevance of climate-induced risks. As shown in Fig. \ref{fig:pilot_overview}, Senigallia is a riverine and coastal city along the Adriatic coastline of the Marche region.Flood risk is mainly relevant for the city center (including the historical area), built on a floodplain (fig. Fig. \ref{fig:pilot_overview} -A).The city suffered from many severe floods. The most recent ones, in May 2014 and September 2022, caused over 5,000 homes flooded, 10,000 people evacuated, and more than 50 casualties and injuries. The economic impact of these floods exceeded 180 million euros. In particular, the Misa River, which crosses the historical center, has masonry embankments vulnerable to floodwaters, while the compact urban fabric, with scarcely permeable surfaces and narrow streets, exacerbates the flood risk. The area of the Foro Annonario, placed in the old town, represents one of the most significant hot-spot in the whole urban fabric (Fig. \ref{fig:pilot_overview} -B), having relevant cultural and architectural value, and hosting  many mass gatherings all over the year. The area is undergoing projects for urban transformation, mainly involving Palazzo Gherardi, a historical multi-storey building, owned by the Municipality, which will become a key public cultural hub. 

\section{RESULTS: FRAMEWORK DEFINITION IN VIEW OF PILOTS USER STORIES AND REQUIREMENTS}
\begin{figure}
    \centering
     \includegraphics[width=\linewidth]{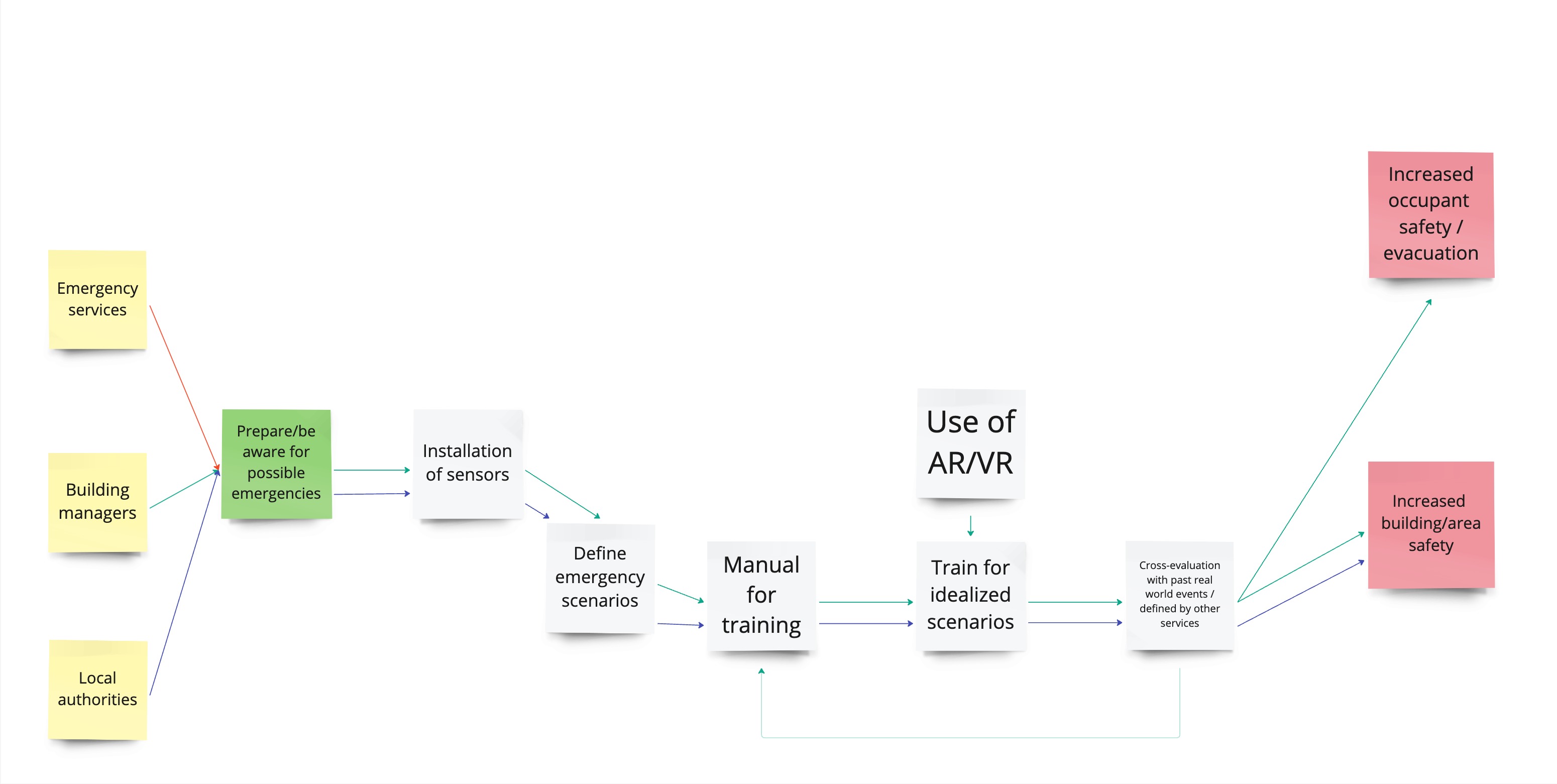}
    \caption{User Flow Diagram, stressing users (in yellow), starting (in red), intermediate (in white) and end (in red) points of the flow.}
    \label{fig:user_flow}
\end{figure}
The whole user flow diagram provided combining the proposal of CLIMRES researchers and the pilot stakeholders' involvement is provided in Fig. \ref{fig:user_flow}. The flow considers three types of users, stressed by yellow labels, and moving from flow starting point (in green), to intermediate actions (in white), up to end points (in red. In particular, representatives and stakeholders from the Municipality of Senigallia allowed the refinement of the overall user flow, and the definition of operational issues to develop the integrated framework starting from functional and non-functional requirements. According to the pilot features as traced in Section II.E, the application is oriented towards flood risk, and mainly comprises the response of individuals to emergency conditions, including the activation and completion of the evacuation process.

The Municipality representatives expressed interested in an XR-based training platform, for immersive and interactive VR/AR technologies. They are also interested in defining a structured training program to increase the emergency preparedness of citizens and first responders (as target groups), by promoting specific activities for vulnerable users exposed to flood risk, including children and elderly. However, end-users should also be able to install the VR/AR solutions on their own devices. This would be part of a "self-training" program, where experience-based metrics help users assess their knowledge and preparedness for flood risk. In this way, the contents could be also available to additional categories of end-users, including tourists and visitors. Based on the aforementioned, tailored instructions, training interactions and safety measures should be adapted to different user groups, ensuring accessibility both online and offline of the XR-based tools, and educational materials should be both accessible online and offline. Finally, due to the specificity of the urban area (including the focus building, Palazzo Gherardi), specific risk-related contents on the platform should be defined in collaboration with first responders' associations, to make it consistent with those that end-users could face in the real-world scenario. 

In view of the user requirements defined above, the XR-based framework is organized according to the following steps.
The first step is \textbf{Preparing the training}. Training programs for the citizens and the first responders will be carried out by involving the Municipality advisors and eventual researchers and experts in flood risk assessment from the CLIMRES project. Activities will be carried out by mainly referring to immersive VR headsets to ensure widespread replication of training content. Nevertheless, AR solutions would be considered to promote in-situ training experiences in the building pilot and surrounding areas. This kind of activity will be mainly oriented towards the Senigallia population, including occupants and users of the focus building, as well as easily reproduced for the whole urban territory (e.g. involving children in schools). Remote training experiences need the creation of accounts to access the educational material, and then a login process to an online training platform. This step makes sure that the training is secure and accessible only to authorized users. In this way, the XR-based training could be also performed by additional categories of end-users by themselves, such as single citizens and tourists. Remote experiences could be mainly oriented towards immersive (e.g. headsets) and non-immersive (e.g. for personal devices) VR solutions \cite{trainingscorgie,vrcalvaresi}.

The second step is \textbf{Selecting a training scenario}. In the XR experience, users could choose from different scenarios that simulate selected flood emergency situations, according to those defined by the Municipality advisors and the flood risk experts from the CLIMRES project. Considering the VR application, a "typological" scenario can be selected to test risk awareness and preparedness in a free-of-context condition, and then move towards pilot-sensitive scenarios related to the Municipality of Senigallia. In this sense, the XR-based framework can directly re-use the VR "typological" scenario in different flood-prone urban scenarios \cite{vrflooddamico}. On the contrary, AR application could be pilot-sensitive, considering  additional and interactive content displayed over the real-world visualization.

The third step is \textbf{Initializing and performing the XR simulation}. Users enter an XR environment that lets them interact with and navigate through the emergency setup, looking for the achievement of the assigned tasks, from a serious gaming perspective. A short introduction to flood risk and training objectives could be first included to improve the engagement levels and provide clear instructions to users. Direct feedback will be provided to the users during the test, depending on the performed actions, through simple metrics, such as those based on health bars or score levels \cite{vrcalvaresi}. In addition, basic monitoring of actions performed in XR could be provided for  final training reports, also including quantitative issues in XR experiences (e.g. decision timing).

The forth step is \textbf{Completing the training and receiving final reports}. At the end of the XR experience, the system evaluates the user's performance based on criteria like performed actions, decision speed and accuracy. Then, the users get final reports focusing on awareness and preparedness levels, and suggestions for improvement. Simple metrics and communication will be deployed to boost the users' knowledge improvement, in the spirit of the Sendai framework recommendation, too \cite{Sendai,trainingcampaing}. Considering training campaigns carried out by Municipality advisors, administrators will have access to a copy of the reports, to review the collected performance data and feedback to update and refine the training scenarios, keeping them effective and up-to-date. In this case, data will be managed in a completely anonymous manner.

The last step is \textbf{Repeating, choosing another scenario or exiting the XR experience}-  Users can either repeat the same scenario to improve their skills or choose a different one. In this case, it could be possible that users, who previously tested its abilities in a "typological" scenario, could then try to face real-world and pilot-sensitive ones. In this case, the final comparison of reports from the two experiences will lead them to obtain additional information on their preparedness level. The same approach could be also adopted by Municipality advisors during in-situ training campaigns.  Once the training sessions are complete, users log out of the system, and the session closes securely.

In view of the above steps configuration, and considering the user flow defined in Fig. \ref{fig:user_flow}, the development of the XR-based framework will comprise the main assumptions, inputs, outputs and challenges traced in Table \ref{tab1}.

Finally, the evaluation of the effectiveness of the XR training solutions and programs developed in the project can be pursued thanks to different metrics, essentially based on \cite{vrcalvaresi,vrflooddamico}: (1) closed-ended self-assessment and  self-efficacy questions before and after the training activities, using likert scale scores; (2) level of involvement and engagement of users in tests; (3) number of errors/wrong emergency behaviors performed in tests, or even number of repetitions of VR experiences, also associated with an health bar during the test. Moreover, written feedback could be collected about additional suggestions on the experience, also to be analyses using advances text mining techniques to detect possible response patterns.

\begin{table}[htbp]
\caption{REFERENCE ELEMENTS FOR XR-BASED FRAMEWORK IN THE PILOT APPLICATION}
\begin{center}
\begin{tabular}{|p{1.5cm}| p{6cm}|}
\hline
\textbf{Main Elements} & \textbf{Specifications}    \\                  
\hline
Assumptions & All \textit{hardware and software} required for the training must be properly installed and operational; Trainers and users should have completed \textit{basic training} on how to use VR/AR equipment, and users should have a basic background on emergency procedures (to be eventually included in the XR application); It is assumed that there will always be internet connectivity for streaming real-time data in case of \textit{online XR experiences}. \\
\hline                                               
Inputs & \textit{User data} (preferences and personal device information); \textit{Storylines} (for serious gaming approaches) and scenario conditions ("typological" and pilot-sensitive); \textit{Emergency plans and response protocols}, also depending on the end-users' typology ("typological" and pilot-sensitive). \\
\hline
Outputs &\textit{ Performance data}: analytics on user performance during training scenarios, highlighting areas of strength and those needing improvement. \textit{User feedback compilation}: Feedback from users providing insights into their experience and the usability of the training modules. \textit{Updated training scenarios}: Revised training modules that incorporate recent feedback and data analysis.    \\        \hline
Challenges & Limited bandwidth might affect the \textit{delivery of real-time XR experience}; \textit{Compatibility issues} with different types of personal devices and operating systems should be checked, orienting the XR tool towards specific platforms and system requirements.\\
\hline                                   
\end{tabular}
\label{tab1}
\end{center}
\end{table}

\section{DISCUSSION AND FINAL REMARKS} 

The integration of Extended Reality (XR) technologies into user risk training within urban built environments can contribute to the increase of community resilience against hazards such as floods. This work proposed a framework to this end, related to citizens, local authorities, and emergency responders as representative end-users of XR tools. By leveraging XR, the training experience can be made immersive and interactive, allowing users to engage with simulated emergency scenarios that reflect real-world conditions, thereby prompting active participation in safety practices, and fostering a culture of preparedness within urban communities. The proposed XR-based framework is adaptable, scalable and suitable for customized application in to different urban contexts, and emphasizes stakeholder involvement in its creation. By focusing on multi-role training, it also bridges gaps between identified user groups interested in risk training actions, while the iterative nature of the XR tests allows for continuous improvement based on feedback and performance data.



The first step in the implementation of the XR-based framework involves the identification and collection of relevant input data traced starting from Table \ref{tab1}, for the pilot of Senigallia, and flood risks. Technical partners will work with local authorities and emergency agencies to gather data on past floods, infrastructure, and demographics. Surveys and focus groups will collect user preferences to ensure the training content meets the needs of identified end-users. 
Then, the user experience design will first involve the definition of specific workflow diagrams and mock-ups for the XR training environment. An intuitive and engaging interface will be pursued, guiding the user journey through training scenarios, highlighting key actions and decisions. Beside immersive training experiences through head-mounted displays (i.e. in training programs by local advisors),
 non-immersive XR solutions, such as web-based simulations, desktop and mobile applications, will be developed for accessibility for users who may not have access to advanced XR hardware and ensure wider inclusivity in the risk preparedness initiative.
 
The pilot application will serve as a testing ground for the XR framework, allowing for real-time feedback and adjustments to improve the training experience. 
This can ensure moving towards replication in other urban contexts, verifying that the framework is adaptable and scalable.
Key aspects of extended application and replication campaign, in the Senigallia case study and in other pilots, can mainly involve: (1) verifying the training reliability and effectiveness of "typological" scenarios in other pilots and replicators, still limited to flood risk; (2)  extending the whole framework and training modules to other urban risks, e.g. earthquakes, wildfires,  extreme weather events; (3) evaluating if immersive and non-immersive XR-based training could lead to awareness, preparedness and learning differences.
In particular, for point (2), research indicates that while immersive experiences can enhance engagement and retention of information, non-immersive environments also play a crucial role in making training accessible to a wider audience, still supplying significant cognitive and emotional responses \cite{vrcalvaresi,vrflooddamico,vr360lovreglio}. 

\end{document}